\documentclass[%
 amsmath,amssymb,
preprint,%
]{revtex4-1}

\usepackage{graphicx}
\usepackage{dcolumn}
\usepackage{bm}

\usepackage[utf8]{inputenc}
\usepackage[T1]{fontenc}
\usepackage{mathptmx}
\usepackage{etoolbox}

\makeatletter
\def\@email#1#2{%
 \endgroup
 \patchcmd{\titleblock@produce}
  {\frontmatter@RRAPformat}
  {\frontmatter@RRAPformat{\produce@RRAP{*#1\href{mailto:#2}{#2}}}\frontmatter@RRAPformat}
  {}{}
}%
\makeatother
\begin{document}


\title{In-situ study of mineral liberation at the onset of fragmentation of a copper ore using X-ray micro-computed tomography}

\author{N. Francois}
\affiliation{ARC Training Centre for M3D Innovation, Research School of
Physics, The Australian National
University, Canberra ACT 2601, Australia}
\email{nicolas.francois@anu.edu.au}
\author{Y. Zhang}
\affiliation{ARC Training Centre for M3D Innovation, Research School of
Physics, The Australian National
University, Canberra ACT 2601, Australia}
\author{R. Henley}
\affiliation{ARC Training Centre for M3D Innovation, Research School of
Physics, The Australian National
University, Canberra ACT 2601, Australia}
\author{L. Knuefing}
\affiliation{ARC Training Centre for M3D Innovation, Research School of
Physics, The Australian National
University, Canberra ACT 2601, Australia}
\author{R. Cruikshank}
\affiliation{ARC Training Centre for M3D Innovation, Research School of
Physics, The Australian National
University, Canberra ACT 2601, Australia}
\author{M. Turner}
\affiliation{ARC Training Centre for M3D Innovation, Research School of
Physics, The Australian National
University, Canberra ACT 2601, Australia}
\author{L. Beeching}
\affiliation{ARC Training Centre for M3D Innovation, Research School of
Physics, The Australian National
University, Canberra ACT 2601, Australia}
\author{A. Limaye}
\affiliation{National Computational Infrastructure, The Australian National University, Canberra ACT 2601, Australia}
\author{A. Kingston}
\affiliation{ARC Training Centre for M3D Innovation, Research School of
Physics, The Australian National
University, Canberra ACT 2601, Australia}
\author{M. Saadatfar}
\affiliation{ARC Training Centre for M3D Innovation, Research School of
Physics, The Australian National
University, Canberra ACT 2601, Australia}
\author{M. Knackstedt}
\affiliation{ARC Training Centre for M3D Innovation, Research School of
Physics, The Australian National
University, Canberra ACT 2601, Australia}

\date{\today}

\begin{abstract}

A better understanding of the relation between ore fragmentation and ore texture is a key to the energy efficient extraction of targeted minerals from low grade ore deposits. In this study, X-ray micro-computed tomography is employed to study mineral liberation during the tensile failure and onset of fragmentation of a copper ore. We present the results of experiments based on a high-pressure instrument enabling micro-mechanical studies to be carried out in-situ (inside a micro-CT scanner). This experimental platform enables mapping in 3D of the evolution of a sample of copper ore during an in-situ fragmentation test. 

The fragmentation occurs quasi-statically via tensile-activated nucleation and growth of multiple cracks producing a complex fracture network. Coupling breakage with microstructural information, we determine quantitatively the impact of ore textural features on fracture patterns and mineral liberation. This information can be compared to the ore mechanical behaviour, in particular to measurements of the deformation energy, the strain deformation field or the stress relaxation response. The copper liberation, fragment size distribution and breakage patterns are statistically characterised and related to two dominant comminution mechanisms which are clearly identified in the sequence of tomographic images. 

Our results show that in-situ micro-CT experiments could inform new studies of ore fragmentation at the laboratory scale and may offer new avenues to address current challenges in the design of efficient comminution processes.

\end{abstract}

\maketitle

\section{\label{sec:level1}INTRODUCTION}

The mining industry uses diverse crushing and grinding methods to reduce lumps of ore into smaller fragments. This process of breaking a coherent ore body into many pieces, called comminution, is used to expose valuable minerals from the rock groundmass and to produce ore-concentrate. Ore comminution represents a substantial part of a mine’s energy consumption. Energy-efficient comminution and optimal mineral extraction are key to the economics and sustainable development of critical mineral mining. Current research efforts to better understand the key factors governing geomaterials comminution reflect the need to rationalise empirical practices. Moreover, the increasing demand for minerals pushes the mining industry to exploit low-grade ore deposits. This trend challenges the usual modelling of comminution efficiency in terms of optimal liberation (resource efficiency) at minimal overgrinding (energy efficiency).

Failure and fragmentation of an ore lump involves a range of complex mechanisms at multiple length scales from microscopic crack openings, to the behaviour of large fractures interacting and producing broad distribution of fragment sizes \cite{Jaeger,Gueguen}. In the context of mining applications, several questions on comminution are the subject of current and intense research efforts \cite{Tavares2021,Hobbs2022}. Typical questions are: What are the mechanisms of fragmentation of an ore with a given texture? Is there a dominant mechanism for a given crushing process? How is the prevalence of a fragmentation mechanism reflected in the final fragment size distribution? How is the coupling of breakage mechanisms with the ore texture heterogeneity impacting the efficient liberation of critical minerals?

Recent advances in X-ray microcomputed tomography (micro-CT) may offer new avenues to address these difficult questions and to inform studies of ore micromechanics and fragmentation at the laboratory scale \cite{Cnudde,Noiriel2015,Viggiani2015}, while providing the basis for development of energy-efficient liberation process streams. Indeed, high resolution X-ray scanners and micro-CT reconstruction techniques enable non-destructive observations of the inner structure of rocks, giving access to three-dimensional information on the crack density, mineral texture and heterogeneous structure of rocks. The capability of micro-CT to provide this structural information from microscopic to centimetre scale also makes it a remarkable tool for studying failure mechanisms in geomaterials \cite{Cnudde,Noiriel2015,Geraud,Ghamgosar2016,Viggiani2015}. Over recent years, CT-based research efforts have focused on the realisation of in-situ mechanical tests, whereby a rock is tested inside the scanner \cite{Geraud,Besuelle2006,Lenoir2007,Ottawa_Saadatfar,Hurley2016,Viggiani2004,Wang,Ando2013,Hurley2018,Hurley2020}. This approach provides direct observations of the evolution of a rock microstructure as it deforms and fails. However, to date, the in-situ study of hard rock failure remains extremely difficult and to our knowledge in-situ tomography has been rarely exploited to study a rock on its path to fragmentation \cite{Parapari,Itai2021}.

Here, we present the results of a model in-situ experiment on mineral liberation during the failure and onset of fragmentation of a low-grade porphyry copper ore. This experiment is concerned with tensile activated rock failure which triggers fragmentation. We analyse in-situ multiple stages of the on-going fragmentation process starting with an initial tensile fracture splitting the ore diametrically, followed by tensile-activated nucleation and growth of multiple fractures producing a complex fracture network. During the mechanical test, high-resolution 3D images are acquired and the quasi-static fragmentation process is mapped sequentially. These 3D images enable tracking of the influence of ore texture on the local fracture mechanism and observing the interaction of multiple growing fractures in this heterogeneous rock. Moreover, tomographic data enables quantification of the copper liberation, the fracture density and the fragment size distribution.

\begin{figure*}
\includegraphics[width=16cm]{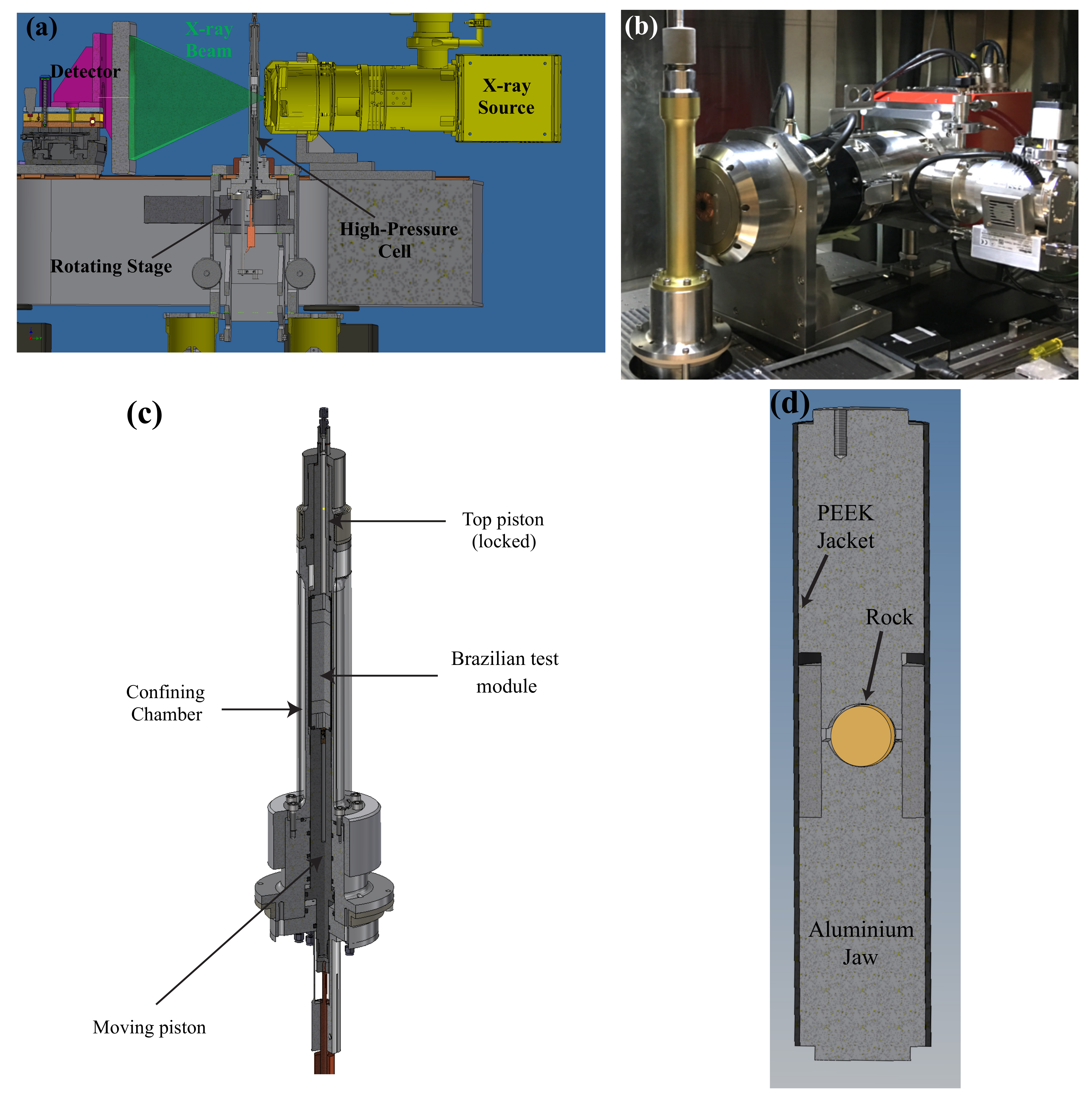}
\caption{\label{Fig1} (a) CAD drawings depicting the XCT-HP apparatus mounted on the high-resolution X-ray micro-tomography platform at the Australian National University. (b) Picture of the instrumental platform showing the confining chamber of the high-pressure cell and the X-ray source. (c) CAD drawing of the main components of the high-pressure cell. (d) Schematics of the crushing-test module. A disc-shaped rock sample is placed in the jaws of the rock-crusher device. The crushing device is placed in the confining chamber of the high-pressure cell.}
\label{Fig1}
\end{figure*}

\section{\label{P1}ORE SAMPLE AND EXPERIMENTAL PLATFORM}

The objective of this experimental study was to explore the mechanical behaviour and fragmentation process of an ore sample subjected to a dominant tensile stress. For this pilot study, a 12 mm diameter 8 mm length sub-core was used. It was extracted from a 45 mm diameter NQ3 drill core from the late Ordovician-early Silurian ($\sim 455 - 435$ $Ma$) Northparkes porphyry copper mine in New South Wales (Australia). The geology of the deposit has been described in \cite{Lickfold2003,Pacey2019}. The deposit is made up of a cluster of intrusion-related porphyry copper ore bodies centred on quartz monzonite porphyry intrusions and each has a copper core within potassium silicate altered intrusive and surrounding rocks \cite{Pacey2019}. The core sample used here is from an altered volcaniclastic breccia that contains clasts of porphyritic andesitic volcanic lava and volcanic sediments ranging from bedded mudstones-siltstones to coarse sandstones and clast rich volcanic breccias.

Central to this study is a unique experimental platform: the X-ray micro-CT high-pressure apparatus (XCT-HP apparatus) which is part of the CTLab based at the Research School of Physics of the Australian National University.

Over the past 15 years, the CTLab has conducted pioneering research in the development of advanced micro-CT techniques \cite{Adrian2014,Kingston2018,Kingston2011,Myers2016} that have been applied to a broad range of problems encountered in geophysics \cite{Knackstedt,Adrian_Review,Feali2012,Qajar,Herring2019,Huang2021,Zhang}, in model granular materials \cite{Aste2005,ANU_PRL2013,ANU_PRL2014,ANU_PRE2015,ANU_NatComm2017} and geomaterials \cite{Ottawa_Saadatfar}, in engineering materials \cite{Saadatfar2012,Flex_Network2014} or more recently of CO2 geological storage in permeable geologic formation \cite{Herring_CO2,Herring_CO2_2}. One of the core strengths of the CTLab is the ability to measure, characterise and quantify 3D physical structures based on the acquisition of high-resolution ($\sim$ 1 $\mu$m), high-quality (high signal to noise ratio) 3D images. The design of the instrument grew out of this research. In the context of geomechanics, this instrument produces very high-quality 3D images which are pivotal to identifying key microscopic displacements and 3D microstructural changes during rock failure and fragmentation.

The XCT-HP apparatus is fully integrated to a cone-beam X-ray micro-CT instrument that was designed and built in-house at the ANU \cite{Francois2022}. Figures 1 (a) and (b) show the main components of this instrumental platform, namely, the high-pressure cell, an X-ray source, an X-ray photon detector and a high precision sample manipulator consisting of a rotation stage and a vertical translation stage. Figure 1 (c) shows the different components of the high-pressure cell. The design of the cell is based on two opposed pistons positioned at the extremities of a cylindrical confining chamber. The top piston is locked in position while the lower piston moves axially and is driven hydraulically by a syringe pump. The syringe pump is connected to a PID controller that enables an accurate control of the axial pressure applied to the moving piston. The pressure in the hydraulic loading line is measured using pressure transducers and simultaneously the piston displacement is measured by a Linear Variable Differential Transformer which provides an additional input to the PID control loop. These various controllers and transducers enable a control of pressure and piston displacement in a $\pm 3.5\times10^{-3}$ MPa and $\pm 0.5$ $\mu$m range respectively.

\begin{figure*}[b]
\includegraphics[width=15cm]{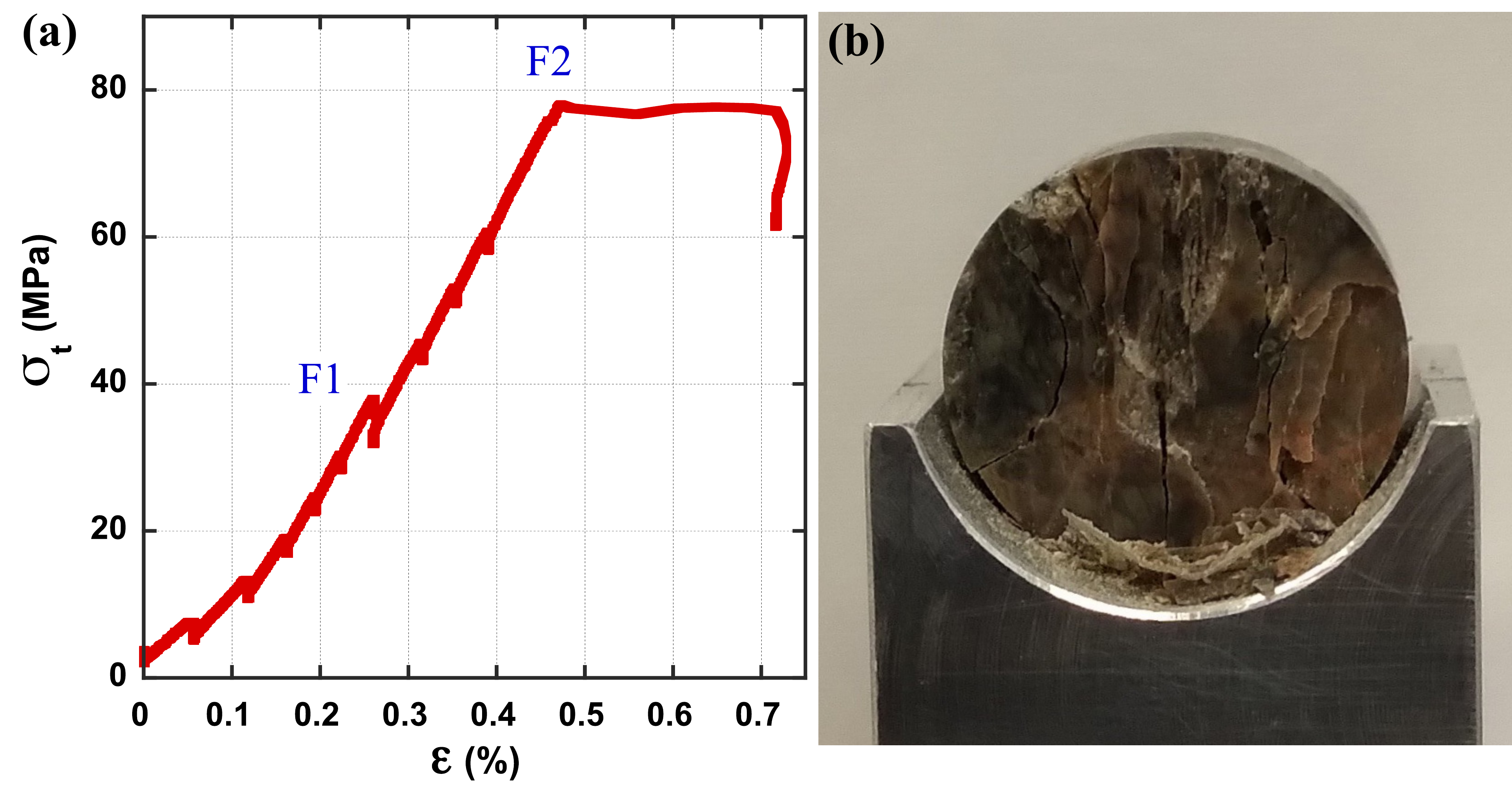}
\caption{\label{Fig2} (a) Stress-strain relation during the in-situ tensile strength test performed on the copper sulphide ore. Indices F1 and F2 indicate the tensile failure and onset of fragmentation. (b) Picture of the fragmented ore.}
\label{Fig2}
\end{figure*}

To perform a tensile activated fragmentation test, we use an in-house designed crushing device. This crusher module is shown in figure 1(d), the loading configuration is based on that of Brazilian tensile tests \cite{BrazTest_Review}. It is composed of a pair of curved Aluminium jaws encapsulated in a cylindrical jacket made of PEEK. This module sits in the confining chamber of the high-pressure apparatus. A disc-shaped ore sample (diameter d = 12 mm, thickness t = 8 mm) is placed in the crusher. As the ore is compressed between the two jaws, a compressive stress is directed along the diameter of the rock while tensile stresses develop in the horizontal direction. The axial force F exerted by the hydraulic piston on the jaws may be converted to a tensile stress $\sigma_t=F/(d \times t)$ \cite{BrazTest_Review}. The strain $\epsilon_t$ is computed as $\epsilon_t=\delta l/d$ where $\delta l$ is the change in the gap distance between the jaws.  Twelve high-resolution tomographic images were acquired at different stages of the mechanical test of the ore sample. The typical scanning time for an image is 17 hours. The image spatial resolution (i.e. the voxel size) is 9 $\mu$m. These images allow sequential quantitative tracking of the quasi-static failure and fragmentation of the ore. The loading protocol between each image consisted of a pressure ramp at a rate of  $1.4\times10^{-3}$ MPa/s followed by a 4-hour long relaxation period at fixed strain value before the image acquisition.

\section{\label{P2}EXPERIMENTAL RESULTS}

Figure 2 (a) shows the stress-strain curve of the copper sulphide ore measured during a 15-day-long experiment. These mechanical measurements can be compared to tomographic sections of the rock obtained at increasing values of the tensile stress (see Figs. 3 (a) and 4 (a)). These two figures clearly show how the rock undergoes successive tensile-activated failure and fragmentation. The images also reveal an interesting fact about the stress-strain curve shown in figure 2 (a). An elastic deformation region seems to exist for a range of strain values extending up to $\epsilon_t=0.47\%$, however a large tensile-activated fracture (F1) occurred at a strain $\epsilon_t=0.26\%$, and a stress of 38 MPa. It was possible to load the ore beyond this failure point, and a linear relation between stress and strain was recovered until the rock sample could not sustain any further increase of the stress. This mechanical event is correlated to the emergence of a dense network of fractures, it occurred at a strain $\epsilon_t \approx 0.47 \%$, and a stress of 78 MPa. Figure 2 (b) shows the fragmented rock after unloading from the high-pressure cell. 

\begin{figure*}
\includegraphics[width=16cm]{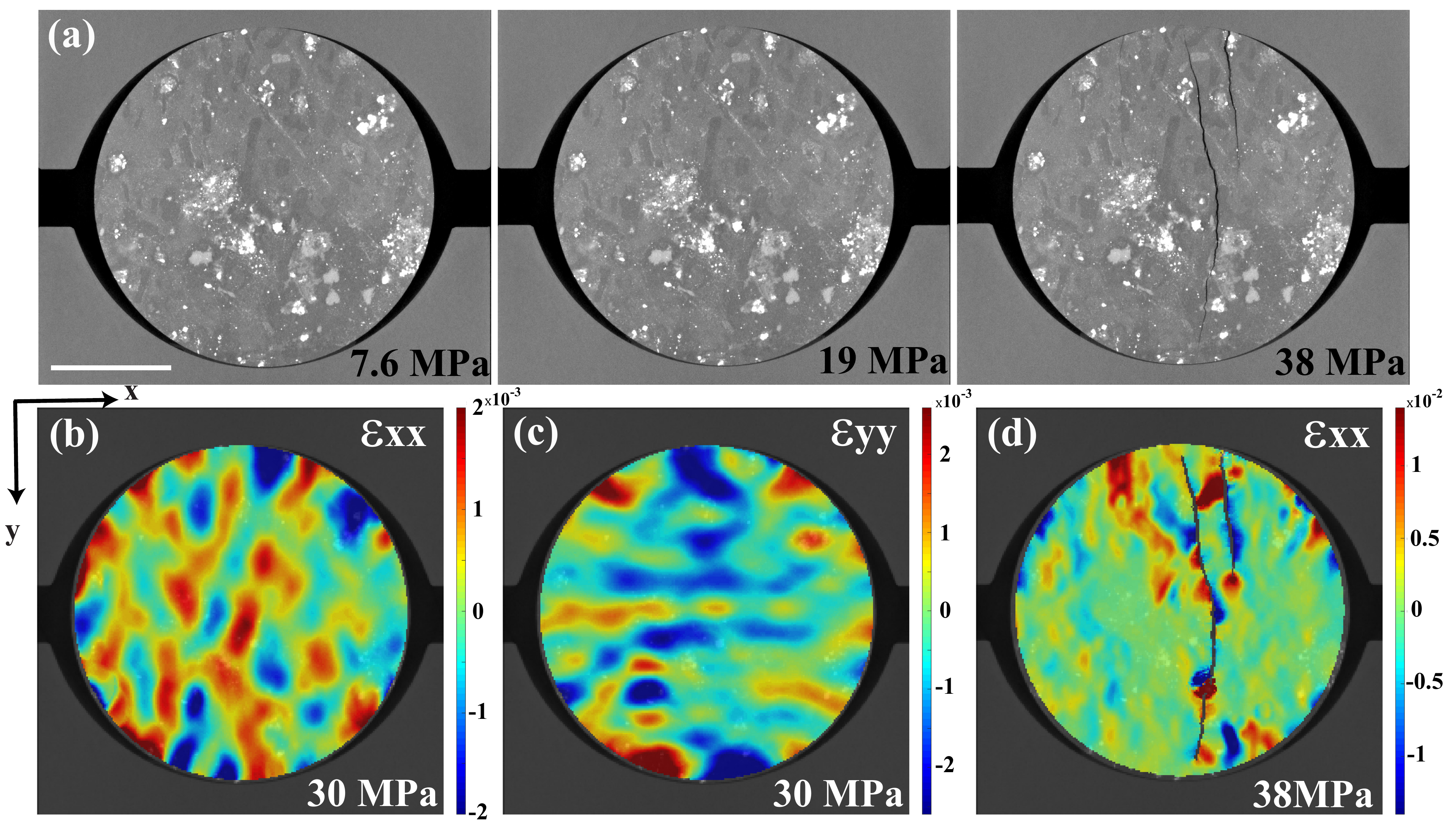}
\caption{\label{Fig3} (a) Midplane tomographic sections of the ore sample at increasing applied load until tensile failure occurs at $\sigma_t=38$ MPa (Scale bar = 4.5 mm) (b-c) Maps of the strain components $\epsilon_{xx}$ and $\epsilon_{yy}$ measured for an increase of $\sigma_t$ from 7.6 MPa to 30 MPa. (d) Strain map of $\epsilon_{xx}$  measured between the stressed state at $\sigma_t=30$ MPa and the tensile failure at $\sigma_t=38$  MPa.}
\label{Fig3}
\end{figure*}

\begin{figure*}
\includegraphics[width=17cm]{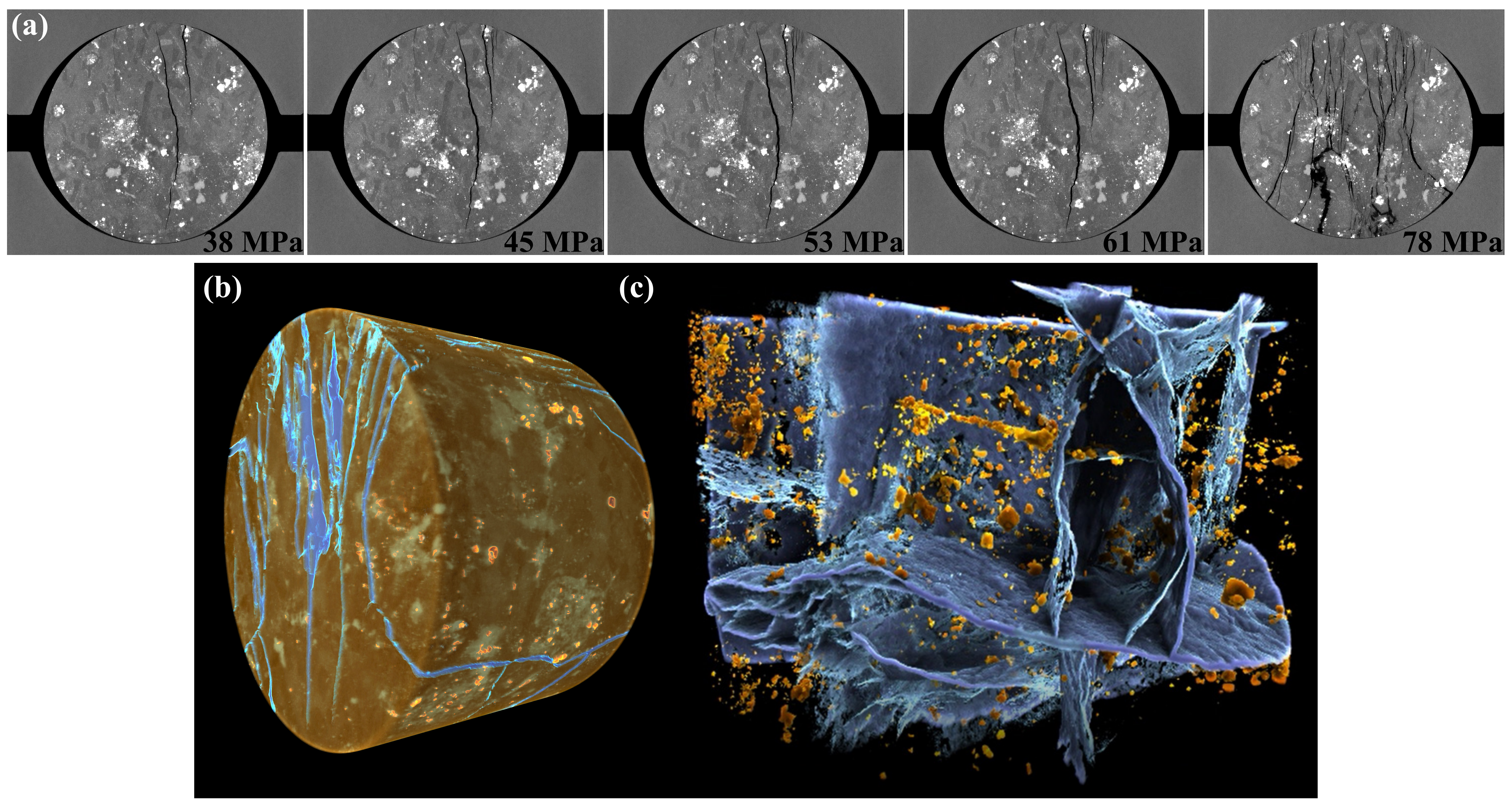}
\caption{\label{Fig4} (a) Midplane tomographic sections showing the ore undergoing successive failure ($\sigma_t=38$ MPa) and fragmentation ($\sigma_t=78$ MPa). (b) 3D reconstruction of the fragmented ore (at $\sigma_t=78$ MPa): fractures (blue) and Cu-rich minerals (orange) are highlighted (rock diameter = 12 mm). (c) 3D visualization of the fracture network and the Cu-rich grains.}
\label{Fig4}
\end{figure*}

The quantitative analysis of the tomographic images provides valuable information on many aspects of the breakage mechanism that leads to the rock fragmentation. For instance, the deformation field of the copper ore is mapped using digital image correlation technique in figure 3 \cite{Lenoir2007}. Local strain amplitudes as low as $5\times10^{-5}$ are measured using a 500-$\mu$m-diam interrogation window on a spatial grid with a mesh size of 50 $\mu$m. In figures 3 (b) and (c), both tensile and compressive strains are observed on the maps of different components of the strain tensor. The spatial distribution of strain is quite different to that computed for homogeneous model rocks \cite{BrazTest_Review}; this difference reflects the complex coupling between the ore heterogeneous structure and the deformation field. The maps in figures 3 (b) and (c) were computed by comparing a reference image with the image acquired at $\sigma_t=30$ MPa just before the tensile failure. The axial strain component $\epsilon_{yy}$ shows an interesting structure with alternative compression and dilation bands along the vertical direction. The transverse strain component $\epsilon_{xx}$ shows the presence of tensile strain hot spots in particular in the central region and the region close to the top jaws. The locations of these hot spots are correlated with the opening of the three large fractures observed at $\sigma_t=38$ MPa (shown in the last panel of figure 3 (a)). Figure 3 (d) shows a map of the strain component $\epsilon_{xx}$ measured at the tensile failure point when the first three fractures opened. As expected, there is a high tensile deformation at the tips of the different fractures. We also note the interesting interaction between the tips of the two central fractures which produces a region with both high tensile and compressive strains. The map also reveals the spatial distribution of compressive and tensile strain along the fracture contours.

The tomographic sections shown in figure 4 (a) illustrate the multiple stages that leads to the onset of ore global fragmentation. The initial large fracture splits the ore diametrically. As stress is further increased, we observe tensile-activated nucleation and growth of multiple fractures producing a complex fracture network. Remarkably the high-pressure apparatus and the PID controller enable us to stabilise the ore sample in a critical mechanical state, in a sense allowing to study quasi-statically the fragmentation process. Figure 4 (b) shows a 3D in-situ visualisation of the ore in its fragmented state. Taking advantage of this digital twin of the fragmented ore, segmentation techniques can be used to visualise the geometry of the fracture network while highlighting the locations of the copper clusters (see figure 4 (c)).

This microstructural information can be compared to different facets of the mechanical response of the ore to tensile stresses. Figure 5 (a) shows the mechanical work done by the crushing jaws computed as $E_w=\int F \delta l$. A sharp change in the evolution of $E_w$ occurs at $\delta l_t = 56$ $\mu$m, marking two different regions. $\delta = \delta l_t $ marks the onset of fragmentation. For $\delta < \delta l_t $, $E_w$ can be interpreted as a deformation energy of the ore sample which accounts for elastic and plastic deformations as well as fracture opening beyond the failure point F1. The ore fracturing occurs at $E_w=0.12$ J and the onset of fragmentation at $E_w=0.42$ J. Beyond the threshold value $\delta l_t$, the ore is yielding and the mechanical response is dominated by the dynamics of the fractures. In this regime, the interpretation of the energy $E_w$ becomes difficult because some large fractures are opening while some other are closing. The closing of a fracture is a highly dissipative process which impacts the ore comminution, this will be discussed below with reference to figures 6 and 7.	

\begin{figure*}
\includegraphics[width=16cm]{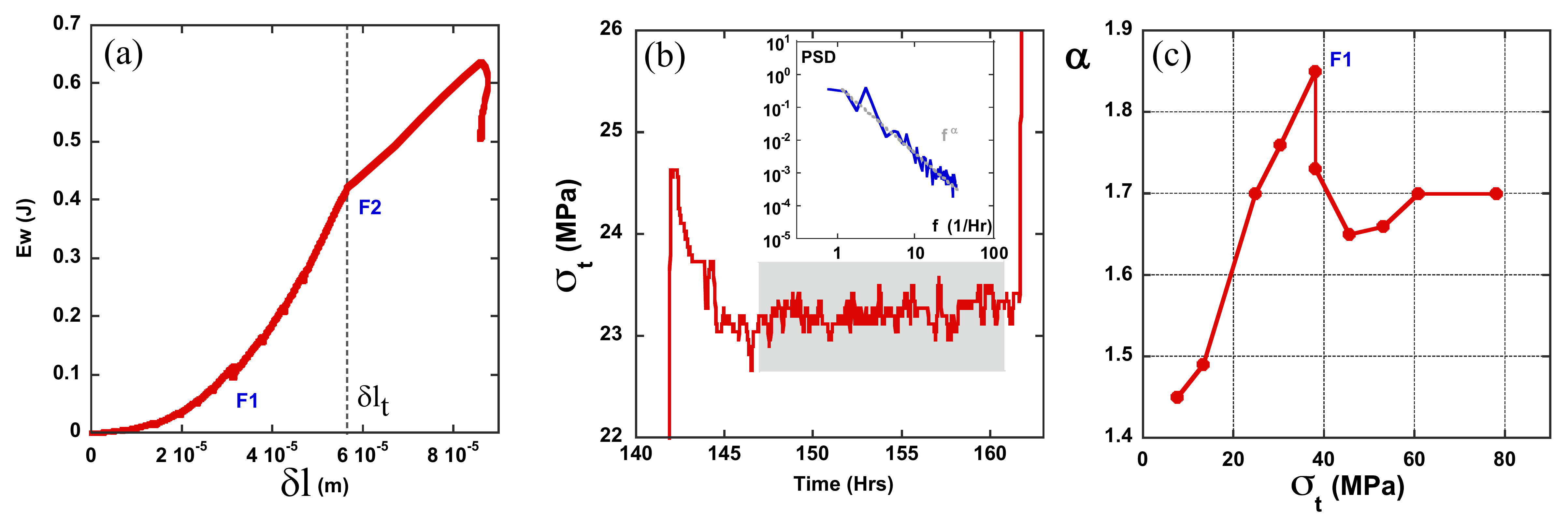}
\caption{\label{Fig5} (a) Mechanical work $E_w$ done by the crushing jaws versus $\delta l$ the reduction of the gap between the jaws. (b) Temporal fluctuations of the tensile stress measured at fixed strain during an image acquisition. The grey box indicates the time domain over which the spectral analysis is carried out. Inset: Typical frequency power spectrum of the stress fluctuations. The grey line shows the scaling law behaviour $f^{\alpha}$. (c) Scaling law exponent $\alpha$ measured at various stages of the mechanical test.}
\label{Fig5}
\end{figure*}

The XCT-HP apparatus enables the study of the mechanical relaxation of the ore at different stages of the fragmentation process.  The transient response to a change in mechanical deformation is an interesting phenomenon in geomaterials \cite{Jaeger}. One typical transient experiment involves deforming a rock and observing the decay in stress over time under a constant strain level. In our experiment, the tensile stress is ramped up to a set value and after a conditioning stage, a tomographic image is acquired at fixed strain $\epsilon_t$. During the 17-hour-long image acquisition, the relaxation of the tensile stress $\sigma_t$ at fixed $\epsilon_t$ is continuously recorded. Figure 5 (b) shows a typical time trace of $\sigma_t$ measured over many hours. The fluctuations of  $\sigma_t$ are expected to be correlated to microstructural modifications occurring in the ore sample. To identify such correlations, we compute the frequency power spectrum of fluctuations of  $\sigma_t$ at different stages of the mechanical test. These power spectra present a broad band with a scaling law behaviour where very low frequencies are excited (see inset Fig. 5 (b)). Figure 5 (c) shows the measurement of the scaling law exponent $\alpha$ at increasing value of the tensile stress. Before fracture, the exponent $\alpha$ increases with the increase of $\sigma_t$, and peaks at the failure point, beyond which the exponent is saturating at a lower value $\alpha=1.7$. These results suggest that the spectral analysis of the mechanical stress fluctuations during stress relaxation could be used to predict the proximity to mechanical failure.

\begin{figure*}[b]
\includegraphics[width=16cm]{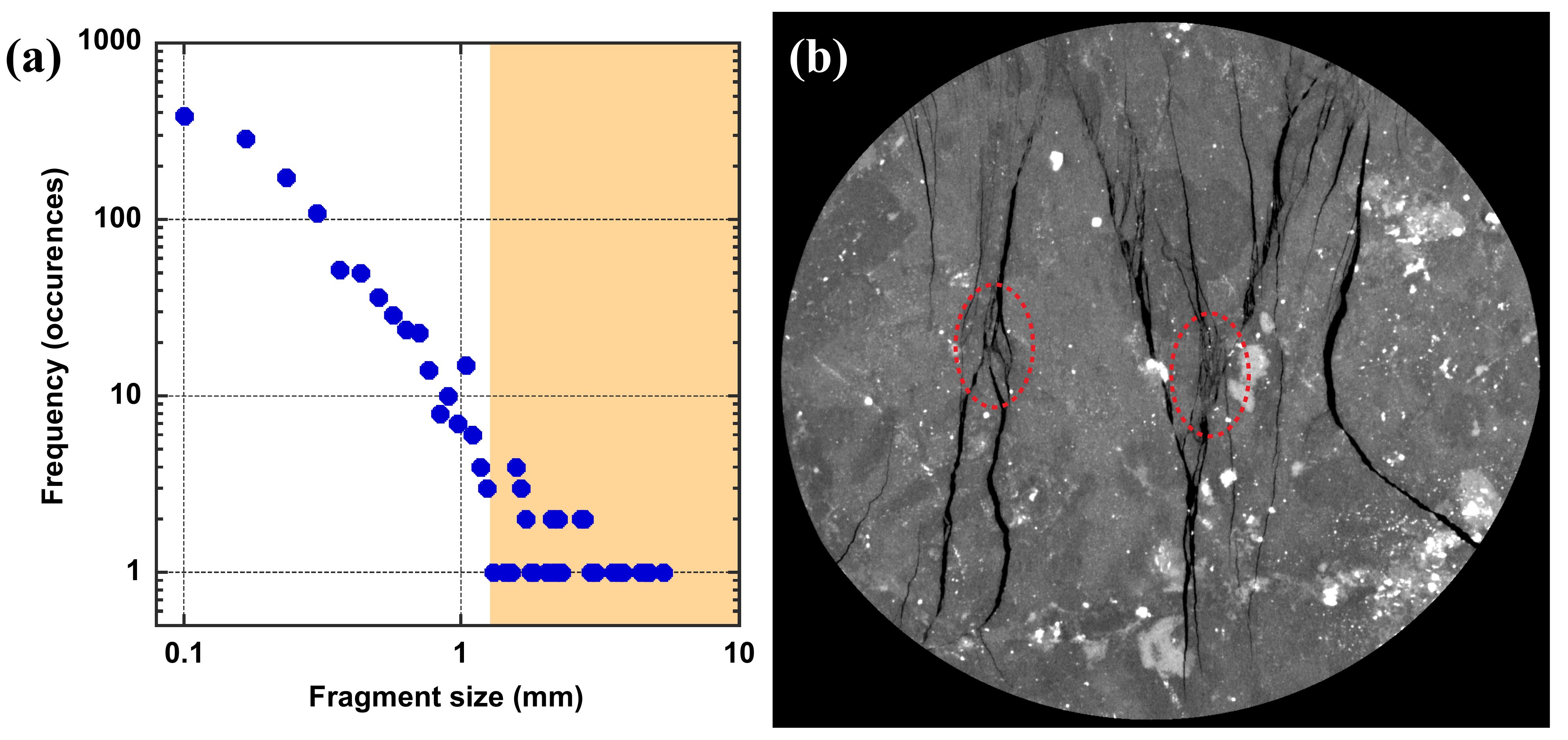}
\caption{\label{Fig6} (a) Fragment size distribution measured after the fragmentation point F2. The fragment size $l$ is computed as $\sqrt[3]{V_f}$ where $V_f$ is the volume of the fragment. The orange area corresponds to the 50 largest fragments for which $l>2$ mm. (b) Tomographic section of the fragmented ore; regions of intense comminution are highlighted by red ellipses. In these regions, the closing of large fractures produces high compression and frictional stresses yielding small size fragments.}
\label{Fig6}
\end{figure*}

Tomographic data also enable precise measurement of two key descriptors of the ore comminution, namely, the fragment size distribution and the damage index. Figure 6 (a) shows the distribution of fragment size as measured at the last stage of the mechanical test. The fragment size $l$ is computed as $\sqrt[3]{V_f}$ where $V_f$ is the volume of the fragment (note that the voxel size is 9 $\mu$m). Though the crushing of the ore is still at its onset, the distribution of fragment size is already quite broad with a mean value at $\bar{l}=250$ $\mu$m. Quantitatively, the 50 largest fragments have a size $l>2$ mm and represent $90 \%$ of the ore volume, while the remaining $10 \%$ of the ore volume is made of more than 1600 fragments. This large population of small fragments arises from two different breakage mechanisms:\\
 1/ as the ore yields beyond the fragmentation point F2, several large fractures are actually closing up and therefore produce high compressive and shear stresses (see regions highlighted by red ellipses in Fig. 6 (b)); the rock-rock friction and compression stresses when the two faces of a fracture come into contact produce small fragments with a broad range of sizes,\\
 2/ other regions yielding many small fragments are located at the two rock-jaw contact points. Indeed, the geometry of the loading induces compressive stresses concentration at the contact point, this stress concentration is known to influence the failure process and to induce local damages \cite{BrazTest_Review}. In these locations, first damages appear after the initial fracture F1. As the loading increases, this process gives rise to dense local fracture networks whose final states are shown in figure 7 (a).\\

We should emphasise two challenges related to the measurements of the fragment size distribution. First, it relies on an accurate detection of the fracture location, such measurements are inherently dependent on the spatial resolution and the contrast level of the tomographic image. In this experiment, the voxel size was $9$ $\mu$m, for future work we plan to use advanced image analysis method based on deep learning to improve the precision of our fracture measurements. Another difficulty is related to the segmentation of the ore image to identify its different fragments. We have used so-called grain partitioning methods which are very efficient in identifying the rounded grains of porous rocks. The application of these methods to fragmented ore is difficult because the ore fragments present a great diversity of size and shape. 

\begin{figure*}
\includegraphics[width=15.5cm]{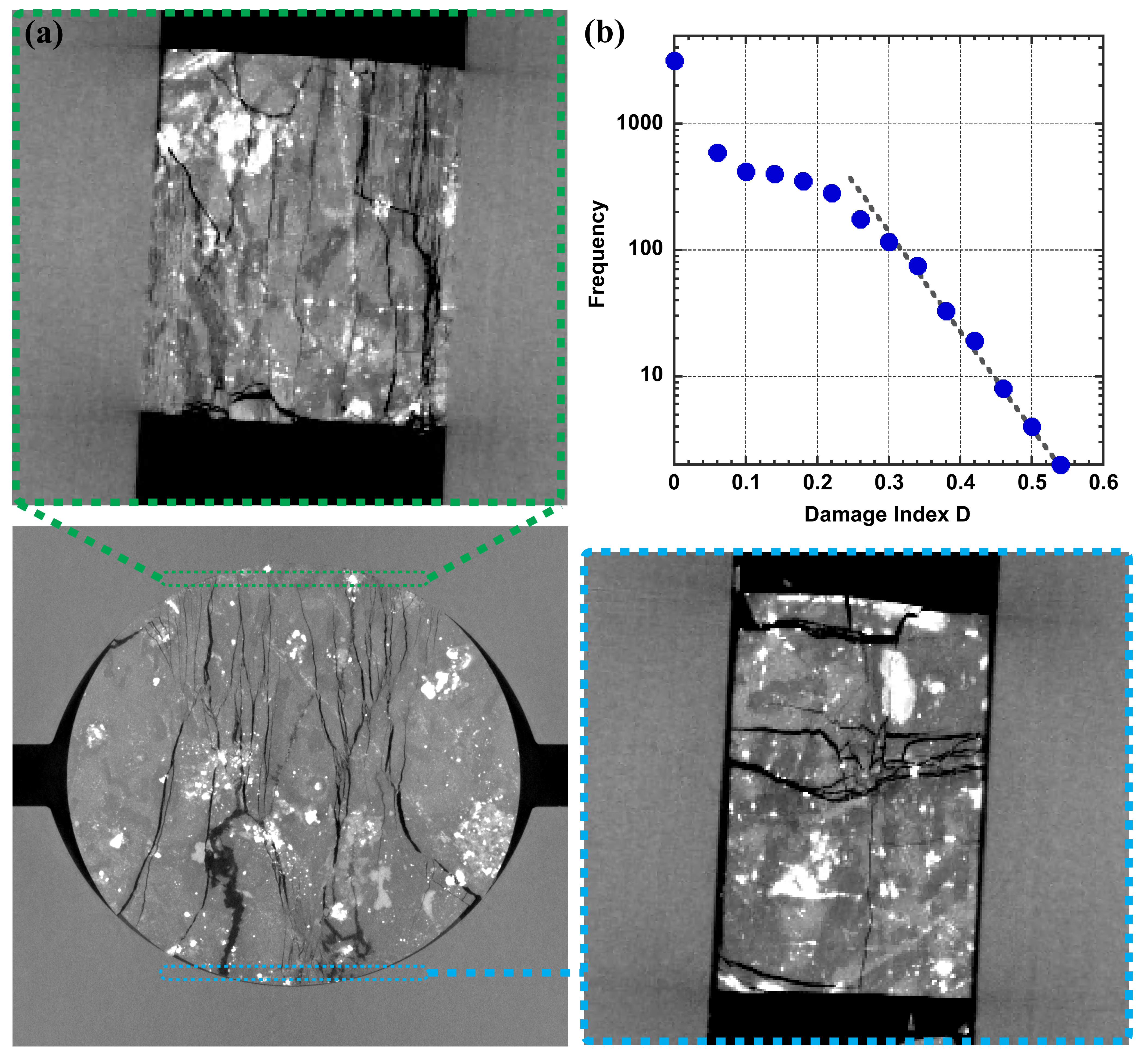}
\caption{\label{Fig7} (a) Vertical and horizontal tomographic sections showing regions of high comminution of the ore. Top and right panels show horizontal tomographic sections close to the contact regions between the rock and the aluminium jaws, where compressive stresses concentration produces a broad range of fragment size. (b) Damage index distribution computed on a 3D spatial grid with a mesh size of 0.5 mm.}
\label{Fig7}
\end{figure*}

To complement this quantitative description of the fragmented ore, we have also measured a damage index $D$. This descriptor is a popular measure of damage introduced by Kachanov \cite{Kachanov1958}. It is based on a continuum mechanics approach to estimate the degree of damage from 2D sections of rocks. In 2D, the definition of the damage index $D$ is given by $D=A_v/A$ where $A$ is the surface area of the rock section and $A_v$ is the area occupied by the cracks and fractures, also named the void area. In this study, we extend this definition to the 3D case, and compute the damage index as $D=V_v/V$ where $V$ is the volume of the ore and $V_v$ is the volume occupied by the fractures. In a sense, $D$ quantifies the fracture density in the ore. When computed on the totality of the ore sample then $D_t=0.072$. This descriptor can also be used to analyse the field of damage. To do so, local values of $D$ were computed on the units of a 3D spatial grid with a mesh size of 0.5 mm. Figure 7 (b) shows the distribution of the local value of $D$ measured by using this approach. This distribution presents a strong peak at $D=0$ corresponding to the undamaged parts of the ore. It also reveals a high probability of observing a damage index in the range $0.1<D<0.2$ that is associated with the fine cracks and the region of high comminution (such as those highlighted by red ellipses in Fig. 6 (b)). Finally, the distribution shows an exponential tail for large values of $D$ that is related to the many fractures that are wide open.

\begin{figure*}
\includegraphics[width=16cm]{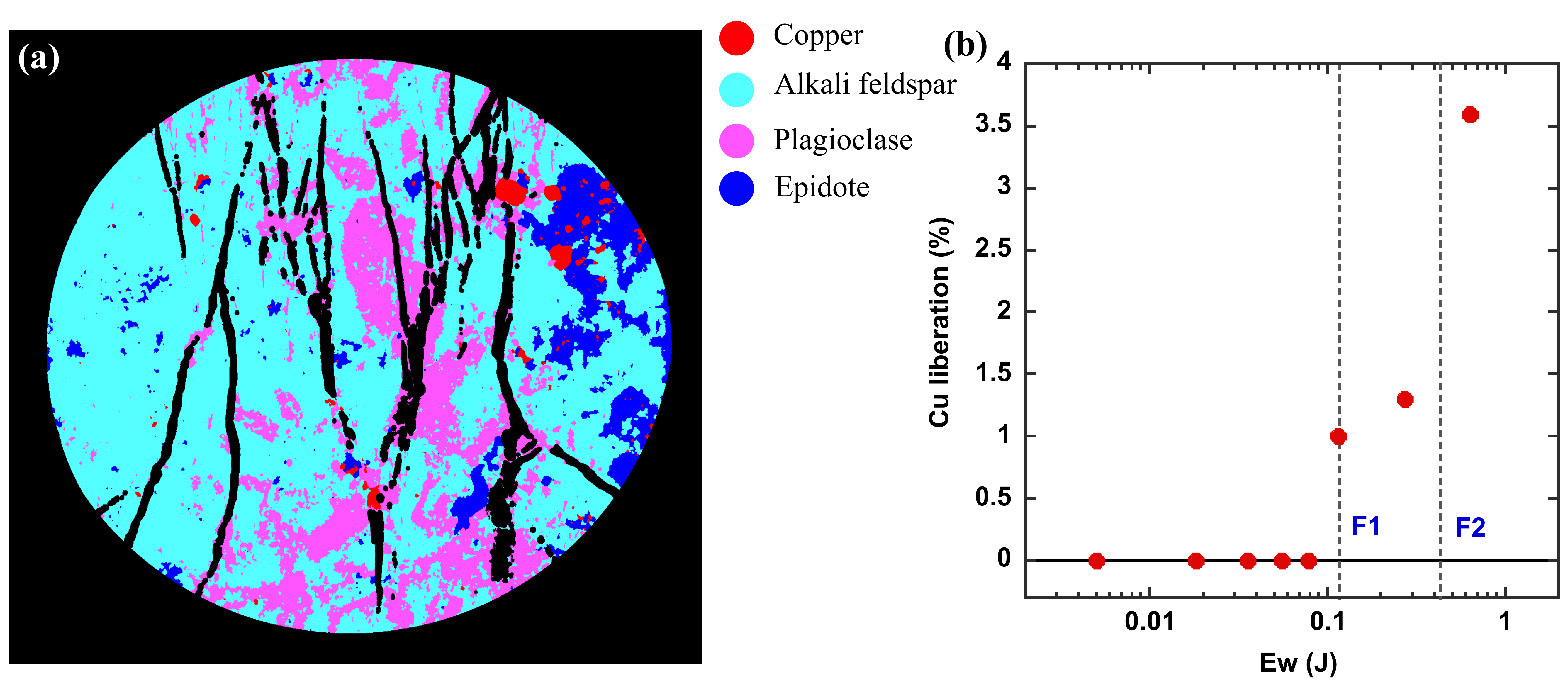}
\caption{\label{Fig8} (a) Texture map of the fragmented ore. The texture map has been computed by combining SEM EDS data with X-ray attenuation values of the tomogram. In this section of the 3D map, the coarse-grained ore texture presents four main phases (Alkali feldspar, Plagioclase, Epidote and Copper sulfides), the fracture location is coloured in black. (b) Evolution of the Copper liberation versus the deformation energy.}
\label{Fig8}
\end{figure*}

A better characterisation of the coupling of breakage statistics (fragments size distribution, local damage index $D$) with texture breakage and critical mineral liberation is key to improving current fragmentation models. In the recent study \cite{Zhang_2022}, we have described a workflow integrating simultaneous 3D mapping of fracture locations and of the dominant mineral phases in an ore sample. Figure 8 (a) shows such a texture map of the fragmented ore at $\sigma_t=78$ MPa. This map has been computed by combining SEM EDS data with the X-ray attenuation values of the tomogram. The ore texture is coarse-grained to present the dominant mineral phases only. Quantitatively the ore is made of: $67.2 \%$ of Alkali feldspar, $24.4 \%$ Plagioclase and $7.1 \%$ Epidote (expressed in volume fraction). Moreover, in our ore sample, there is $1.3 \%$ of copper sulphides (bornite and chalcopyrite). We emphasize that such estimations are not trivial, because it relies on the use of many experimental techniques, each bringing its own challenges as discussed in \cite{Zhang_2022}. The fracturing of each phase can be clearly identified and quantified as the creation of new surface area compared to the unfractured state. In the fragmented state, $56 \%$ of the new surface area was produced in the feldspar phase, $39 \%$ in the plagioclase, $4.3 \%$ in the epidote and $0.7 \%$ in the copper. 
The degree of liberation of copper (the proportion of copper exposed by fracturing) is shown in Figure 8 (b) as a function of the deformation energy $E_w$.  Two regimes of liberation are observed: after the rock fracturing (event F1), approximately $1 \%$ of the copper sulphides is freed from the gangue and this fraction increases weakly until the rock is fragmented (event F2). In the final fragmented state, the copper liberation increases substantially to $3.6 \%$. This evolution is consistent with three main features of the ore texture and breakage behaviour: 1/ the copper sulphides grains are mainly embedded in the epidote and feldspar phases (with the highest contact preference with the epidote), 2/ the plagioclase and feldspar are the most damaged phases during the fracturing and subsequent fragmentation, 3/ the epidote is only substantially damaged beyond the onset of fragmentation. The latter process explains the increase in the liberation of copper sulphides observed beyond the event F2. In this experiment, we only investigated the onset of ore fragmentation which explains the relatively low ($< 5 \%$) degree of liberation of the copper grains. On-going work focusses on measuring copper liberation at more advanced stages of the fragmentation process.

\section{\label{P3} Conclusion}

In summary, this experiment tackles a tensile-activated rock fracturing which triggers the fragmentation of a porphyry copper ore. The results presented here show how high-resolution X-ray micro-CT has the potential to link breakage fundamentals and ore texture features such that this combined information may assist in the design of optimal processes for critical mineral liberation.		
This study also shed light on the potential of in-situ mechanical testing for empirical modelling of comminution processes as used in the mining industry. In-situ data enable simultaneous quantification of mineral liberation, mechanical energy dissipation, fragment size distribution and of the coupling of local fracturing with the ore texture. Such data are needed to inform the modelling of optimal liberation (resource efficiency) at minimal overgrinding (energy efficiency).
In terms of rock mechanics, this example of an in-situ fragmentation test, based on a simple loading configuration, opens up interesting opportunities to model in the laboratory complex breakage mechanisms. In this study, the focus was on comminution based on rock-rock friction induced by fracture closing and on fragment production at rock-metal compression contacts. 
On-going work includes in-situ investigation of more advanced stages of the fragmentation process with a focus on the relation between fragment size distribution and mineral liberation.

\begin{acknowledgments} 
This work was supported by the Australian Research Council Training Centre M3D (ARC IC180100008).
\end{acknowledgments}

\end{document}